\documentclass[twocolumn]{article}
\usepackage{graphics}

\title{Modelling (001) surfaces of II-VI semiconductors}
\author{M. Ahr\footnotemark[1] 
\and M. Biehl \and T. Volkmann \and
Institut f\"{u}r Theoretische Physik und Astrophysik \\ 
Julius-Maximilians-Universit\"{a}t W\"{u}rzburg
\\ Am Hubland, 97074 W\"{u}rzburg, Germany}

\begin{document}
\twocolumn[
\maketitle
\begin{abstract}
First, we present a two-dimensional lattice gas 
model with anisotropic interactions which explains the experimentally observed 
transition from a dominant $c(2\times 2)$ ordering of the CdTe(001) surface to a local 
$(2\times 1)$ arrangement of the Cd atoms as an equilibrium phase transition. Its analysis 
by means of transfer-matrix and Monte Carlo techniques shows that the 
small energy difference of the competing reconstructions determines to a large extent 
the nature of the different phases. Then, this lattice gas is extended to a
model of a three-dimensional crystal which qualitatively reproduces many of the 
characteristic features of CdTe 
which have been observed during sublimation and atomic layer epitaxy. \vspace{2em}
\end{abstract}
]
\footnotetext[1]{Corresponding author. Phone: +49 (0) 931 888 4908
Fax: +49 (0) 931 888 5141 E-mail: ahr@physik.uni-wuerzburg.de}
Within the last years, potential applications of electronic devices based
on II-VI semiconductors have inspired basic research concerning for instance
the surface reconstructions of these materials. 
For CdTe, a fairly complete qualitative phase diagram of $(001)$ surfaces 
has been obtained \cite{ct97}. An 
understanding of the interplay of these reconstructions with techniques like 
molecular beam epitaxy (MBE) or atomic layer epitaxy (ALE) is desirable both for 
technological applications and for the theory of crystal growth. 
From the theoretical point of view, the case of CdTe is especially interesting, 
since it exhibits a phase transition which involves competing reconstructions 
at typical MBE growth temperatures. 
While there have been a few models of MBE, especially for GaAs which take into 
account the effects of surface reconstructions (e.g. \cite{kps01}), phase transitions between 
different reconstructions have not been considered in growth models, so far. 

The zinc-blende lattice of CdTe consists of alternating layers of Cd and Te 
atoms the positions of which lie on regular square lattices. 
Under vacuum, the CdTe(001) surface is always Cd terminated. The surface is characterized by 
vacancy structures where the maximum Cd coverage $\rho_{Cd}$ is $1/2$. 
At low temperature, one finds a $c(2\times 2)$ reconstruction, where the Cd atoms 
arrange in a checkerboard pattern. 
Above a critical temperature $T_c = 543 K$\cite{ntss00} the surface is dominated by a 
local $(2\times 1)$ 
ordering where the Cd atoms arrange in rows along the $[110]$ direction
which alternate with rows of vacancies \cite{ct97,tdbev94,ntss00}. Electron diffraction 
experiments have shown that 
there is a high degree of disorder in the $(2\times 1)$ phase \cite{ntss00}. 
This suggests an interpretation of the reordering as an  order-disorder phase transition. 
An external Cd flux stabilizes the $c(2\times 2)$ reconstruction at $T > T_c$. 
If a Te flux is applied, one finds surfaces terminated by Te dimers or trimers \cite{ct97,tdbev94}. 

We first investigate whether the above mentioned transition can be explained within
the framework of thermodynamic equilibrium. 
In a simplifying manner we treat the terminating Cd layer of the surface 
as a two-dimensional square lattice gas with sites either being occupied by a Cd 
atom or empty \cite{baksv00}.
Chemical bonds and the influence of the underlying crystal are accounted for 
by effective pairwise anisotropic interactions. An infinite repulsion excludes nearest 
neighbour (NN) pairs in $ [1\overline{1}0]$-direction 
(denoted as $y$-direction in the following). This can be justified from 
considerations like the electron counting rule \cite{ct97}. In
$[110]$(x)-direction an attractive interaction $\epsilon_x < 0$ 
favors the occupation of NN sites. A 
competing attractive interaction $\epsilon_d < 0$ between diagonal neighbors (NNN) tends to stabilize
the $c(2\times2)$ arrangement. The total energy of the system then reads
\begin{eqnarray}
 H &=& \epsilon_d n_{Cd,d} + \epsilon_x n_{Cd,x} - \mu n_{Cd},
\end{eqnarray}
where $n_{Cd}$, $n_{Cd,d}$ and $n_{Cd,x}$ denote the total number of Cd atoms,
NNN and NN pairs in x-direction, respectively. The (effective) chemical potential
$\mu$ controls the mean Cd coverage $\rho_{Cd}= \left< n_{x,y} \right>$. By choosing 
$\epsilon_d =-1$ the energy scale is fixed. 
In agreement with DFT calculations \cite{gffh99}, $\epsilon_x$ is chosen such that 
perfect $c(2\times 2)$ and $(2 \times 1)$ arrangements 
are nearly degenerate, the groundstate being  a $c(2\times 2)$ reconstruction. 

Besides the coverage $\rho_{Cd}$ the system is characterized by 
the two-point correlations $C_{Cd}^{x} = n_{Cd,x}/n_{Cd}$ and 
$C_{Cd}^{d} = n_{Cd,d}/n_{Cd}$ which determine the fraction of 
Cd atoms incorporated in locally $(2\times 1)$- or $c(2\times 2)$- ordered regions. 

\begin{figure}
\resizebox{\columnwidth}{!}{\includegraphics{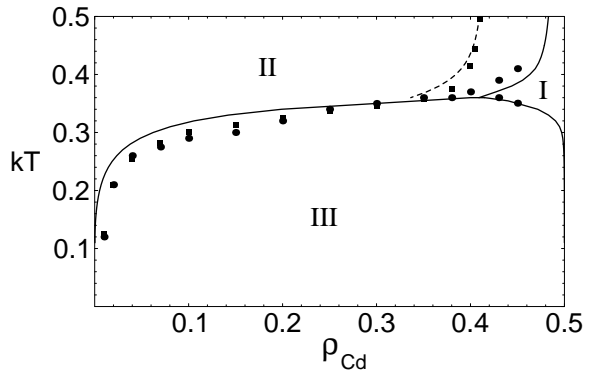}}
\caption{Phase diagram of the 2D lattice gas with $\epsilon_x=-1.90$. 
Solid lines and circles represent
the phase boundaries, obtained from TM and MC. Right of the dashed line (squares) the 
$c(2\times 2)$ structure is prevalent and vice versa. \label{zwei}} \end{figure}

Through transfer-matrix (TM) calculations and Monte Carlo (MC) simulations 
at constant $\rho_{Cd}$ we obtain an estimate 
of the phase boundaries in the temperature-coverage plane (see fig.
\ref{zwei}). At low temperatures (III), a $c(2\times 2)$ ordered phase with 
$\rho_{Cd}\approx 1/2$ coexists  with a disordered phase of low coverage. At higher
temperatures, the system becomes homogeneously disordered (II) or ordered (I) depending on the
coverage. For a range of coverages $\rho_{Cd}$ the transition from (III) to (II) is 
accompanied by a sudden decrease of $C_{Cd}^{d}$ and a simultaneous increase of 
$C_{Cd}^{x}$, meaning that for these
values of $\rho_{Cd}$ the $(2\times 1)$-structure is {\it{locally}} prevailing in the disordered
regime (the area left of the dashed line (or squares) in fig. \ref{zwei}). 
With decreasing $|\epsilon_x|$ the tricritical point, i.e. the point in the 
$\rho_{Cd}$-$T$-diagram
where (I),(II) and (III) meet, shifts to smaller coverage and higher temperature.
Even more so does the line which separates $c(2\times 2)$ from $(2\times 1)$ prevalence. 
This might explain, why a transition from $c(2\times 2)$ to $(2\times 1)$ prevalence has 
not been found in ZnSe, so far. There, the energy difference between perfect 
$c(2\times 2)$ and $(2 \times 1)$ reconstructions is indeed greater than for CdTe 
\cite{gffh99}.  

In order to investigate the evolution of the morphology of the surface in 
non-equilibrium processes like growth and sublimation, we extend the planar 
lattice gas to a model of a three-dimensional crystal. In a previous publication, 
this has been done for a simple cubic crystal which consists of alternating layers 
of the two atomic species \cite{ab01}. 
In this paper, we present the next step towards a realistic modelling of CdTe by simulating 
the correct zinc-blende lattice structure. 
Besides the Cd-Te bond (binding energy $\epsilon_c$) we consider interactions between atoms
on NN and NNN sites in one layer of the crystal. 
The basic idea here is to model the interactions of Cd atoms on the surface with the 
anisotropic interactions $\epsilon_x$, $\epsilon_d$ and a hardcore repulsion in the 
$y$-direction while there is an 
isotropic attraction between Te atoms ($\epsilon_{Te, b}$) and Cd atoms  
inside the bulk of the crystal ($\epsilon_{Cd, b}$). 

Due to the topology of the 
zinc-blende lattice the condition of {\em ergodicity} can not be fulfilled 
within the limits of 
a solid-on-solid (SOS) model, where each atom must be bound to at least two atoms of the 
opposite species in the layer below. 
This can be achieved by considering the formation of Te trimers which have been 
observed experimentally. 
Additionally, we consider Te atoms which have only one bond to a neighbouring
Cd atom. 
In the following, Te atoms in these configurations
will be denoted as Te$^*$ atoms.
Te$^*$ atoms neutralize the hardcore repulsion between Cd atoms.
Otherwise, the deposition of Te on a Cd terminated surface with a $c(2 \times 2)$ 
reconstruction would be impossible.  

As a simplification, we assume that point defects are not incorporated into the crystal. 
Then, the bulk can be described in a SOS manner, while the Te$^*$ atoms 
are considered seperately. 
In summary, the Hamiltonian of the model is 
\begin{eqnarray}
H &=& \epsilon_c n_c + \epsilon_{Cd,b} n_{Cd,b} + \epsilon_{Te,b} + \epsilon_h 
n_{Cd,h} 
\nonumber
\\ &+& \epsilon_x n_{Cd,x} + \epsilon_d n_{Cd,d} + \epsilon^* n^*  
\end{eqnarray}
which is a function of the numbers of Cd-Te bonds ($n_c$), Te$^*$ atoms ($n^*$), 
NN and NNN pairs of surface Cd atoms ($n_{Cd,x}$, $n_{Cd,d}$), 
NN pairs of Cd atoms in the bulk and Te atoms ($n_{Cd,b}$, $n_{Te,b}$) and surface 
Cd atoms next to incorporated Cd atoms ($n_{Cd,h}$). 
Pairs of surface Cd atoms which are not bound to Te$^*$ atoms on neighbour sites in the 
$y$-direction are forbidden.

\begin{figure}
\resizebox{\columnwidth}{!}{\rotatebox{270}{\includegraphics{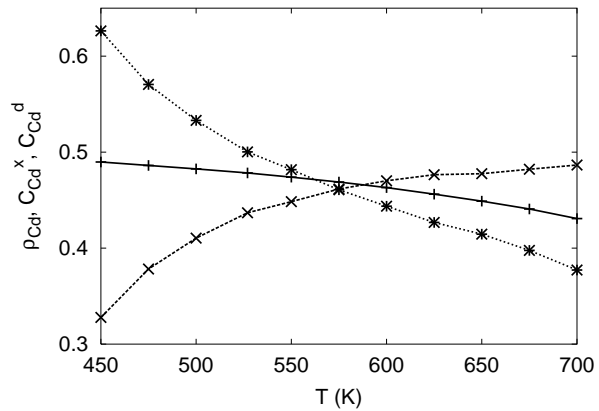}}}
\caption{$\rho_{Cd}$(+) and the correlations $C_{Cd}^{x}$($\times$), $C_{Cd}^{d}$(*) as functions of 
temperature in the stationary state of sublimation. The system size was
$64 \times 64$. \label{drei}}
\end{figure} 
Besides the deposition of adatoms we consider desorption, 
diffusion to NN and NNN sites and diffusion across steps. 
These are Arrhenius-activated processes with rates $\nu \exp[-E/(k T)]$. Desorption requires 
an activation energy $E = \Delta H$. We have chosen Metropolis-like energy barriers 
$E = \mathrm{max} \{B, B + \Delta H \}$ for diffusion processes. 
There is a barrier $B = B_0$ for the diffusion of atoms with two bonds and a 
barrier $B^*$ for the diffusion of Te$^*$ atoms. 

Unfortunately, the available experimental data are not sufficient for a systematic 
fit of the model parameters and there are no values of diffusion barriers available from
ab initio or semi-empirical calculations. Therefore, the choice of the parameter set 
is partly based on estimates and physical reasoning.
In the following, we present results that have been obtained with the parameter set 
$\epsilon_d = -0.086 eV$, $\epsilon_x = -0.168 eV$, $\epsilon_{Cd, b} = \epsilon_{Te, b} 
= -0.069 eV$, $\epsilon_h = 0$, $\epsilon_c = -0.43 eV$, $\epsilon^* = -0.26 eV$, $B_0 = 0.78 eV$,  
$B^* = 0.17 eV$ and $\nu = 10^{12}/s$. 

Experiments have revealed a high density of steps on CdTe surfaces with a 
preferential orientation in the $(100)$ direction.
This leads to a dominant contribution of step flow 
sublimation. Therefore, we have simulated sublimation on vicinal surfaces 
with steps parallel to the $[100]$ direction at a distance of $16$ lattice 
constants. 

Our model reproduces the observation of Cd terminated surfaces in vacuum. 
A Te terminated surface decays in a temperature-dependent time intervall $\tau(T)$. 
 The corresponding desorption probability $p(T) \sim 1/\tau(T)$ follows an 
Arrhenius law with an activation energy of $1.25 \pm 0.02 eV$. 
After this onset, we obtain a stationary 
state where Cd and Te evaporate 
stoichiometrically. Then, the temperature dependence of the sublimation rate follows an Arrhenius law
with a higher activation energy of $1.437 \pm 0.003 eV$. 
Experimentally, activation energies of $1.54 eV$ \cite{ntss00} and $0.9eV$ \cite{tdbev94} have been 
found for stoichiometric step flow sublimation and the decay of a Te terminated surface. 

\begin{figure}
\resizebox{\columnwidth}{!}{\rotatebox{270}{\includegraphics{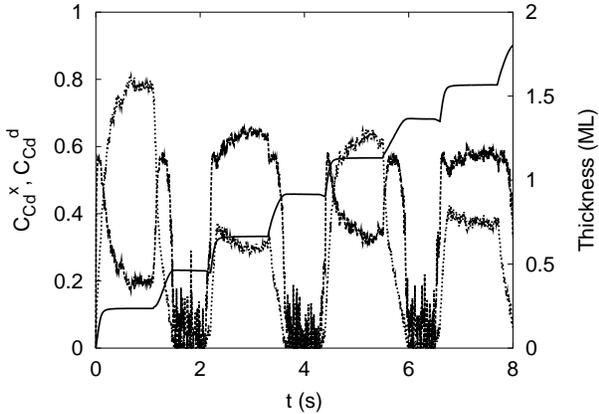}}}
\caption{Correlations $C_{Cd}^{x}$ (dashed), $C_{Cd}^{d}$ (dotted) and thickness of the deposited layer 
(solid) in ALE as a 
function of time. The pulse time was 1s, followed by 0.1s of dead time, the system size was $128 \times 128$. \label{vier}}
\end{figure}

Figure \ref{drei} shows $\rho_{Cd}$ and the correlations $C_{Cd}^{x}$, $C_{Cd}^{d}$ in 
the stationary state as functions of $T$. In the investigated temperature interval, 
$\rho_{Cd}$ decreases only slightly from 
$\rho_{Cd} = 0.49$ at $T = 450K$ to $\rho_{Cd} = 0.43$ at $T = 700K$. 
At low temperature, the surface is dominated by a $c(2 \times 2)$ arrangement of the Cd
atoms ($C_{Cd}^{d} > C_{Cd}^{x}$). On the contrary, at high temperature the Cd atoms 
tend to arrange in the rows of the $(2 \times 1)$ reconstruction ($C_{Cd}^{d} < C_{Cd}^{x}$). 
The crossover between both regimes is at $T_c = 570K$. This is the counterpart of the phase 
transition observed in the 2D lattice gas model. 
However, an investigation of $C_{Cd}^{x}$, $C_{Cd}^{d}$ in 
the 2D lattice gas at the $\rho_{Cd}$ measured in the 3D model shows, that the non-equilibrium 
conditions of sublimation increase the dominance of $(2\times 1)$ over $c(2 \times 2)$ in the 
high temperature regime. 

ALE provides a scenario where our model can be investigated in a situation present in technical 
applications. The basic idea is to obtain self-regulated layer-by-layer growth by alternate deposition
of pure Cd and Te. 
Figure \ref{vier} shows the evolution of the thickness of the deposited layer and the correlations
$C_{Cd}^{x}$, $C_{Cd}^{d}$  at $T = 500K < T_c$. The initial state was a flat Te terminated surface. 
In the first phase, a Cd flux of $5 ML/s$ is applied, in the second phase Te at the same 
rate. At the end of the first phase, the surface is Cd terminated with a $c(2\times 2)$ reconstruction. 
In the Te phase, Te terminated islands form which, at the end of the cycle, cover $\approx 50 \%$ 
of the surface. 
This is due to the fact that only half of the Cd atoms needed for a closed monolayer are present. 
Note, that the decrease of $\rho_{Cd}$ at the beginning of the Te cycle leads to an 
increase of $C_{Cd}^{x}$, which is consistent with the phase diagram (fig. 1) of the 2D lattice gas. 
In the following Cd phase, the islands are covered with Cd. Surprisingly, now the reconstruction
is $(2 \times 1)$. This can be understood from the fact that the lattice gas interactions are 
present only between particles in the same layer. Thus, the island edges impose open boundary 
conditions to the lattice gas of Cd atoms on the island. In constrast to a $c(2\times 2)$ reconstructed 
domain, a $(2 \times 1)$ terminated island can reduce the energy of its boundary by elongating in
$x$-direction. Since the ground state energies of both structures are nearly degenerate, the 
formation of $(2\times 1)$ may thus reduce the surface free energy. In the following Te phase, 
Cd and Te atoms diffuse into the gaps between the islands, which leads to the formation of 
a closed, Te covered monolayer again.
These observations closely resemble ideas suggested in \cite{alepaper} to explain 
experimental observations. 

In summary, we have presented a model of CdTe(001) surfaces which reproduces several experimental 
observations in a semi-quantitative way. In the future, we will extend this model by considering 
the dimerization of surface Te atoms and the fact that Te is adsorbed and desorbed as Te$_2$. 
In cooperation with an experimental group, we will try a more quantitative determination of 
an appropriate parameter set.

\end{document}